\documentclass[aps,preprint,amsmath,amssymb]{revtex4}
\usepackage{graphicx}

\begin{document}

\title{\bf
 T violation in $B\to K \phi \phi $ decays }
\date{\today}
\author{\bf  Chuan-Hung Chen$^{a}$\footnote{Email:
physchen@mail.ncku.edu.tw} and Chao-Qiang
Geng$^{b}$\footnote{Email: geng@phys.nthu.edu.tw} }

\affiliation{$^{a}$Department of Physics, National Cheng-Kung
University, Tainan 701, Taiwan \\
$^{b}$Department of Physics, National Tsing-Hua University,
Hsinchu  300, Taiwan}

\begin{abstract}
We present the general form of the decay width angular
distributions with T-odd terms in $B\to K \phi \phi$ decays. We
concentrate on the T violating effects by considering various
possible T-odd momentum correlations. In a generic class of CP
violating new physics interactions, we illustrate that the T
violating effect
  could be more than 10\%.

\end{abstract}
\maketitle

One of main goals in $B$ factories is to study  CP
violation (CPV), which was first discovered  in the kaon system
\cite{CCFT} 40 years ago. Recently, Belle \cite{BelleCP} and Babar \cite {BabarCP} Collaborations have also
confirmed that the CP symmetry is not conserved in the $B$ system.
Although the standard model (SM) with three generations could provide
a CP violating phase in the Yukawa sector \cite{CKM}, our
knowledge on the origin of CPV is still unclear because it
is known that the same CP violating phase cannot explain the
observed asymmetry of matter and antimatter. That is, searching a
new CP violating source is  one of the most important issues in $B$
factories.

As known that the CP-odd quantities which are directly related to
the CP violating phases can be defined as the decay-rate
difference in a pair of CP conjugate decays. Such kind of the CPV
will depend on two phases, one is the weak CP violating phase and
the other is the strong CP conserved phase. In addition, one can
also define some other useful observables by the momentum
correlations. In B physics,  T-odd
triple-product correlations, denoted by $\vec{p}_i\cdot
\vec{\varepsilon}^*_{1}\times \vec{\varepsilon}^*_{2}$, 
 in the two-body $B\to V_{1} V_{2}$ decays,  have been studied in Refs. \cite{BVV1,BVV2},
 where $\vec{p}_i\ (\vec{\varepsilon}_i)$ is the three-momentum (polarization) of the vector meson $V_i$. 
The experimental searches for such correlations are in progress at $B$ factories
\cite{Todd}. For three-body B decays, there are many possible types of  correlations and 
the simplest ones are
the triple correlations of $\vec{s}\cdot ( \vec{p}_{i}\times
\vec{p}_{j})$ \cite{CG-PRD66}, where $\vec{s}$ is the spin carried
by one of outgoing particles and $\vec{p}_{i}$ and $\vec{p}_{j}$
denote any two independent momentum vectors. Clearly, the triple
momentum correlations are T-odd observables since they change sign
under the time reversal ($T$) transformation of $t\rightarrow -t$.
In terms of the CPT invariant theorem, T violation (TV) implies
CPV. Therefore, by studying of T-odd observables, it could help us
to understand the origin of CPV. We note that these observables
of the triple momentum correlations do not require strong
phases. In this paper, we study the possibility to observe  T
violating effects in the three-body decays at B factories.

Recently, Belle \cite{Belle-PRL} has observed the  decay branching ratios (BRs) of $B^{\pm}\to K^{\pm} \phi
\phi$ are large, which are $(2.6\pm 0.3)\times 10^{-6}$ with the $\phi \phi$ invariant
mass below $2.85$ GeV. By the naive analysis, the decaying mode is
dictated by the process $b\to s \bar{s} s$ at the quark level,
arising
 from the one-loop penguin mechanism. 
 In Ref.~\cite{Hazumi}, it has been shown that the direct CP-odd
observable associated with a new CP violating phase in the decays
could have an
excess of $5$ standard deviations with $10^9$ B mesons. Since the
final states of $B^{\pm}\to K^{\pm} \phi \phi$ involve two vector
mesons which  provide more degrees of freedom due to
 spins, many triple momentum correlations can be constructed.
 It is interesting to investigate the possibility of
observing these T-odd observables due to CPV in these decays.
We note that Datta and London \cite{BVV2} have considered the unique triple momentum correlation with new physics in $B\to \phi K^*$ which is also related to the process of $b\to s s \bar{s}$. 
However, the three-body decays of  $B^{\pm}\to K^{\pm} \phi \phi$ 
 contain more T-odd observables in which new physics involved can be different from that in  $B\to \phi K^*$ and thus our study provides
  alternative ways to search for
T violating effects.

Since the process of  $b\to s s \bar{s}$ is dominated by
loop effects, for simplicity, the corresponding effective
interactions are given by
\begin{equation}
H_{{\rm eff}}={\frac{G_{F}}{\sqrt{2}}}V_{t}
\left[a_{1}O_{1}+a_{2}O_{2}+a_{3}O_{3}+a_{4}O_{4}\right] \;,
\label{heff}
\end{equation}
with the Cabibbo-Kobayashi-Maskawa (CKM) matrix elements
$V_{t}=V_{ts}^{*}V_{tb}$ and the operators
\begin{eqnarray}
O_{1}&=&(\bar{s}b)_{V-A}(\bar{s}s)_{V-A}\;,\;\;\;
\;O_{2}=(\bar{s}_{\alpha}b_{\beta})_{V-A}(\bar{s}_{\beta}s_{\alpha})_{V-A}\;,
\nonumber \\
O_{3}&=&(\bar{s}b)_{V-A}(\bar{s}s)_{V+A}\;,\;\;\;
\;O_{4}=(\bar{s}_{\alpha}b_{\beta})_{V-A}(\bar{s}_{\beta}s_{\alpha})_{V+A}\;,
\end{eqnarray}
where
$\alpha$ and $\beta$ are the color indices and the notations
$(\bar{q}q^{\prime})_{V \mp A}$ stand for the currents $\bar{q}
\gamma_{\mu}(1\mp \gamma_5)q^{\prime}$.
In general, there also
exists a right-handed current $(\bar{s}b)_{V+A}$ associated with
$b$-quark. However, due to that the corresponding transition matrix element $\langle
K| (\bar{s}b)_{V+A}|B\rangle$ involves only the vector current,
 the
contributions from this kind of interactions can be included in
 Eq.
(\ref{heff}) straightforwardly. Moreover, for  the
Wilson coefficients $a_i$ in Eq. (\ref{heff}), the following combinations
\begin{eqnarray}
a^{eff}_{1}&=&a_1+\frac{a_{2}}{N_{c}}, \ \ \
a^{eff}_{2}=a_2+\frac{a_{1}}{N_{c}}, \nonumber \\
a^{eff}_{3}&=&a_3+\frac{a_{4}}{N_{c}}, \ \ \
a^{eff}_{4}=a_4+\frac{a_{3}}{N_{c}},
\end{eqnarray}
with the color factor $N_{c}$ are more useful.
It is known that due to the nonperturbative effects, it is
difficult to deal with the exclusive nonleptonic decays precisely.
In the heavy quark limit, since the particles could be energetic
in three-body $B$ decays, accordingly if we could just
concentrate on all final state particles  in the energetic
region, the leading effect will be factorizable parts and  those
effects from nonfactorizable parts will be subleading. In $B\to K
\phi \phi$ decays,  the region of the
$\phi\phi$ invariant mass measured at Belle is less than the mass of $\eta_{c}$.
That is, both $\phi$ mesons are approximately leaving $B$
 collinear. Then, in the $B$ rest frame, the whole system looks
like a two-body decay. Therefore, outgoing particles are all
energetic. Hence, we assume that the factorization parts are
dominant.
In terms of the factorization assumption, the relevant hadronic
transition matrix elements can be parametrized as
\begin{eqnarray}
\langle K(p_{3}) | \bar{b}\; \gamma_{\mu} (1-\gamma_{5})s |
B(p_{B})\rangle&=& f_{+}(Q^{2})P_{\mu}+\frac{P\cdot
Q}{Q^{2}}Q_{\mu}(f_{0}(Q^{2})-f_{+}(Q^{2})), \label{cb} \nonumber \\
             \langle \phi(\epsilon_{1},p_{1}) \phi(\epsilon_{2},p_{2}) |
\bar{s}\; \gamma_{\mu}\;s | 0\rangle&=& \Big[ \epsilon^{*}_{1}
\cdot \epsilon^{*}_{2} A_{1} + {\epsilon^{*}_{1}\cdot Q\,
\epsilon^{*}_{2}\cdot Q \over Q^2 }A_{2}\Big] (p_{1}+p_{2})_{\mu}
\nonumber \\
&&+ \Big[ \epsilon^{*}_{1} \cdot \epsilon^{*}_{2} B_{1} +
{\epsilon^{*}_{1}\cdot Q\, \epsilon^{*}_{2}\cdot Q \over Q^2}B_{2}\Big] (p_{1}-p_{2})_{\mu} \nonumber \\
&& + C_{1} \epsilon^{*}_{1}\cdot Q \epsilon_{2\mu}+  C_{2}
\epsilon^{*}_{2}\cdot Q \epsilon_{1\mu}, \label{ovc} \\
                \langle \phi(\epsilon_{1},p_{1}) \phi(\epsilon_{2},p_{2}) |
\bar{s}\; \gamma_{\mu}\gamma_{5}\;s |
0\rangle&=&i\varepsilon_{\mu\nu\rho\sigma}
\epsilon_2^{\nu\ast}p^\rho_1p^\sigma_2 (\epsilon_1^\ast\cdot
p_2){D_1 \over m^{2}_{\phi}}
  +i\varepsilon_{\mu\nu\rho\sigma}
\epsilon_1^{\nu\ast}p^\rho_2p^\sigma_1  (\epsilon_2^\ast\cdot p_1){D_2\over m^{2}_{\phi}} \nonumber \\
&&-i \epsilon_{\mu \nu \rho \sigma}\epsilon^{*\nu}_{1}
\epsilon^{*\rho}_{2}\Big( E (p_{1}+p_{2})^{\sigma}+
F(p_{1}-p_{2})^{\sigma}\Big), \label{oavc}
\end{eqnarray}
where $\epsilon_{1(2)}$ denote the polarization vectors of the
$\phi$ mesons, $P=p_{B}+p_{3}$ and $Q=p_{B}-p_{3}=p_{1}+p_{2}$.
The functions $A, B, C, D, E$ and $F$ are the relevant form
factors and functions of $Q^2$. For simplicity, we neglect to
show
their explicit $Q^2$ dependences. Using the equation of motion,
we get
\begin{eqnarray*}
\langle V_{1}(\epsilon_{1},p_{1}) V_{2}(\epsilon_{2},p_{2}) |
\bar{s}\; \not{Q} \;s | 0\rangle&=& (m_{s}-m_{s}) \langle
V_{1}(\epsilon_{1},p_{1}) V_{2}(\epsilon_{2},p_{2}) | \bar{s}\; s
| 0\rangle=0 \;,\nonumber \\
\langle V_{1}(\epsilon_{1},p_{1}) V_{2}(\epsilon_{2},p_{2}) |
\bar{s}\; (\not{p}_{1}-\not{p}_{2})\gamma_{5}\;s | 0\rangle&=&-i E
\varepsilon_{\mu \nu \rho \sigma} (p_{1}-p_{2})^{\mu}
\epsilon^{*\nu}_{1} \epsilon^{*\rho}_{2}
(p_{1}+p_{2})^{\sigma}=0\; ,
\end{eqnarray*}
which imply that  $A_{1}=A_{2}=0$, $C_{1}=-C_{2}$, $D_{1}=D_{2}$
and $E=0$. Hence, Eqs. (\ref{ovc}) and (\ref{oavc}) may be
simplified to
\begin{eqnarray}
\langle \phi(\epsilon_{1},p_{1}) \phi(\epsilon_{2},p_{2}) |
\bar{s}\; \gamma_{\mu}\;s | 0\rangle&=&\Big[ \epsilon^{*}_{1}
\cdot \epsilon^{*}_{2} B_{1} + \epsilon^{*}_{1}\cdot Q\,
\epsilon^{*}_{2}\cdot Q {B_{2} \over Q^2}\Big] (p_{1}-p_{2})_{\mu} \nonumber \\
&&+ C \Big[\epsilon^{*}_{1}\cdot Q \epsilon_{2\mu}
-\epsilon^{*}_{2}\cdot Q \epsilon_{1\mu}
\Big] \label{vc} \\
\langle \phi(\epsilon_{1},p_{1}) \phi(\epsilon_{2},p_{2}) |
\bar{s}\; \gamma_{\mu}\gamma_{5}\;s |
0\rangle&=&i\frac{D}{m^{2}_{\phi}}\Big[(\epsilon_1^\ast\cdot p_2)
\varepsilon_{\mu\nu\rho\sigma}
\epsilon_2^{\nu\ast}p^\rho_1p^\sigma_2
  +i(\epsilon_2^\ast\cdot
p_1)\varepsilon_{\mu\nu\rho\sigma}
\epsilon_1^{\nu\ast}p^\rho_2p^\sigma_1 \Big] \nonumber \\
&&-i F \varepsilon_{\mu \nu \rho \sigma}\epsilon^{*\nu}_{1}
\epsilon^{*\rho}_{2}(p_{1}-p_{2})^{\sigma}. \label{avc}
\end{eqnarray}
In addition, according to the Fierz transformation, the four-Fermi
interaction $(V-A)\otimes (V+A)$ can be transformed to
$(S-P)\otimes (S+P)$. Hence, the matrix elements associated scalar
and pseudoscalar currents can be obtained via equation of motion
to be
\begin{eqnarray}
\langle K(p_{3}) |\bar{b}\; s| B\rangle &=& -\frac{P\cdot
Q}{m_{b}-m_{s}}f_{0}(Q^{2}),\nonumber \\ 
      \langle \phi(\epsilon_{1},p_{1}) \phi(\epsilon_{2},p_{2}) | \bar{s}\; s |
0\rangle&=& \frac{Q^2-(2m_{\phi})^2}{2m_{s}} \epsilon^*_{1}\cdot
\epsilon^*_{2} B_{1} \nonumber \\
                &&+ \frac{\epsilon^*_1\cdot Q
\epsilon^*_2\cdot Q}{2m_s}\left(
\left(1-\frac{(2m_{\phi})^2}{Q^2} \right)B_{2}-2C \right), \nonumber \\
               \langle \phi(\epsilon_{1},p_{1}) \phi(\epsilon_{2},p_{2}) |
\bar{s}\gamma_{5} s | 0\rangle&=&
i\frac{F}{m_{s}}\varepsilon_{\mu\nu\rho\sigma} \epsilon^{*\mu}_{1}
\epsilon^{*\nu}_{2}p^{\rho}_{1}p^{\sigma}_{2}. \label{sc}
\end{eqnarray}
By combining the results of Eqs. (\ref{cb}), (\ref{vc}), (\ref{avc})
and (\ref{sc}), the transition matrix element for $B\to K \phi\phi
$ is expressed by
 \begin{eqnarray}
{\cal M}&=& \frac{G_{F}}{\sqrt{2}}V_{ts}V^{*}_{tb}\left\{
\left(m_{1}\epsilon^{*}_{1} \cdot \epsilon^{*}_{2}+{m_{2}\over Q^2
} \epsilon^{*}_{1}\cdot Q\, \epsilon^{*}_{2}\cdot Q
\right)p_{B}\cdot(p_{1}-p_{2})+im_{3}\Big[{\epsilon^{*}_{2}\cdot Q
\over
m^{2}_{\phi}}\varepsilon_{\mu\nu\rho\sigma}\epsilon^{*\mu}_{1}p^{\nu}_{2}p^{\rho}_{1}p^{\sigma}_{B}
\right. \nonumber \\ && +(1\leftrightarrow2)\Big] +im_{4}
\varepsilon_{\mu\nu\rho\sigma}\epsilon^{*\mu}_{1}
\epsilon^{*\nu}_{2}(p_1-p_2)^{\rho}p^{\sigma}_{B}+im_{5}
\varepsilon_{\mu\nu\rho\sigma}\epsilon^{*\mu}_{1}
\epsilon^{*\nu}_{2} p^{\rho}_{1} p^{\sigma}_{2} \Big\},
\label{amp}
\end{eqnarray}
where various components are defined as
\begin{eqnarray}
m_{1}&=& \frac{m_{11}+m_{12}\cos\theta}{p_{B}\cdot (p_1-p_2)}=
B_{1}f_{0}
\frac{c^{eff}_{3}}{r_{s}}\frac{Q^2-(2m_{\phi})^2}{p_{B}\cdot
(p_1-p_2)}+\frac{4|\vec{p}_{B}| |\vec{p}_{1}|}{p_{B}\cdot (p_1-p_2)}B_{1}c^{eff}_{1}f_{+}\cos\theta,\nonumber \\
m_2&=&\frac{m_{21}+m_{22}\cos\theta}{p_{B}\cdot (p_1-p_2)}= Z
f_{0} \frac{c^{eff}_{3}}{r_{s}}\frac{Q^2}{p_{B}\cdot
(p_{1}-p_{2})}+\frac{4|\vec{p}_{B}| |\vec{p}_{1}|}{p_{B}\cdot
(p_1-p_2)}
 \frac{Zc^{eff}_{1}f_+}{1-\frac{(2m_{\phi})^2}{Q^2}}\cos\theta, \nonumber \\
m_{3 }&=&-2c^{eff}_{2}f_{+}D, \ \ \ m_{4}=2c^{eff}_{2}f_{+}F,\nonumber \\
m_{5}&=&2c^{eff}_{2}\left(\frac{m^{2}_{B}}{Q^2}(f_0-f_+)-f_+
\right)F -2\frac{c^{eff}_{3}}{r_{s}}f_{0}F, 
\label{mij}
\end{eqnarray}
with 
\begin{eqnarray}
c^{eff}_{1}&=&a^{eff}_3+a^{eff}_1+a^{eff}_2, \ \ \
c^{eff}_{2}=a^{eff}_3-a^{eff}_1-a^{eff}_2, \ \ \
c^{eff}_3=a^{eff}_4,\nonumber\\
Z &=& \left(1-(2m_{\phi})^2/Q^2 \right)B_{2}-2C\,.
\end{eqnarray}

In order to get the spectrum with CP and T violating effects,
we choose the relevant coordinates of  momenta and
polarizations in the rest frame of $Q^2$  as
\begin{eqnarray}
Q&=&(\sqrt{Q^2},0,0,0),\ \
E_{B}=\frac{m^{2}_{B}+Q^2}{2\sqrt{Q^2}}, \ \
|\vec{p}_{B}|=|\vec{p}_{K}|=E_{K}=\frac{m^{2}_{B}-Q^2}{2\sqrt{Q^2}},
\nonumber \\
    p_{{1(2)}}&=&(E_{\phi},\pm p_{\phi}\sin\theta,0, \pm
    p_{\phi}\cos\theta),\ \ E_{\phi}=\frac{\sqrt{Q^2}}{2}, \ \
    p_{\phi}=\sqrt{E^{2}_{\phi}-m^{2}_{\phi}}, \nonumber \\
    \epsilon_{1(2)L}&=&\frac{1}{m_{\phi}} (p_{\phi}, \pm E_{\phi} \sin\theta,0 , \pm E_{\phi}
    \cos\theta), \ \ \epsilon_{1T}(\pm)=\frac{1}{\sqrt{2}} (0, \cos\theta, \pm i,
    -\sin\theta), \nonumber \\
          \epsilon_{2T}(\pm)&=&\frac{1}{\sqrt{2}} (0, \cos\theta, \mp i,
    -\sin\theta),\label{coord}
\end{eqnarray}
where $\theta$ stands for the polar angle of the $\phi$ meson. From
Eqs. (\ref{amp}) and (\ref{coord}),
the
differential decay rate for $B\to K \phi \phi $ as a function of
$Q^2$ is given by
\begin{eqnarray}
\frac{d\Gamma}{dQ^2 }&=&{|V_{tb}V_{ts}|^{2} G^{2}_{F} \over 2^{10}
\pi^3 m_{B}} \left(1-\frac{Q^2}{m^2_B}\right)\sqrt{1-\frac{(2
m_{\phi})^{2}}{Q^2}}\left\{ 2\left[|m_{11}|^2
+\frac{2}{3}|m_{12}|^2 \right]e_{11}\right.
\nonumber \\
&& +2\left[|m_{21}|^2 +\frac{2}{3}|m_{22}|^2 \right]e_{22}+ 2
\left[2 Re(m_{11} m^{*}_{21})+\frac{2}{3} Re(m_{12} m^{*}_{22})
\right]e_{12} \nonumber \\
&&\left.+\left( |m_{3}|^2 e_{33} + |m_{4}|^2
e_{44}\right)+2Re(m_{3}m^{*}_{4})e_{34}+2|m_{5}|^{2} e_{55}
+4Re(m_{4}m^{*}_{5}) e_{45} \right\},
\end{eqnarray}
where
\begin{eqnarray*}
e_{11}&=&2+\frac{(p_1 \cdot p_2)^2}{m^{4}_{\phi}}, \ \ \
e_{22}=\left(\frac{m^{2}_{\phi}}{Q^2} \right)^2 \left(1-\frac{(p_1
\cdot p_2)^2}{m^{4}_{\phi}}\right)^2, \nonumber \\
e_{12}&=&-\frac{p_{12}}{Q^2}\left(1-\frac{p^{2}_{12}}{m_{\phi}^4}
\right), \ \ \ e_{33}=\frac{4}{3}
\frac{8\kappa}{(2m_{\phi})^2}\left(1-
\frac{p_{12}^2}{m^{4}_{\phi}}\right), \nonumber \\
e_{44}&=&
m_B^4\left(1+\frac{Q^2}{m^{2}_{B}}\right)^2\left(1-\frac{(2m_{\phi})^2}{Q^2}\right)
     -\frac{4}{3}\frac{8\kappa}{(2m_{\phi})^2}, \nonumber \\ 
e_{34}&=& \frac{4}{3}\frac{8
\kappa}{(2m_{\phi})^2}\left(1-\frac{p_{12}}{m_{\phi}^2}\right), \
\ \ e_{55}=2p_{12}^2-2m_{\phi}^4, \nonumber \\
e_{45}&=&-m^{4}_{B}\left(\frac{m_{\phi}^2}{m^{2}_{B}}-\frac{p_{12}}{m^{2}_{B}}\right)\left(1+\frac{Q^2}{m^{2}_{B}}\right),
\nonumber \\
\kappa&=&- m^4_B\frac{Q^2}{16}\left(1-\frac{Q^2}{m^2_B}\right)^2
\left(1-(2m_{\phi})^2/Q^2\right),
\nonumber \\
p_{12}&=&(Q^2-2m_{\phi}^2)/2
\end{eqnarray*}
and $m_{ij}$ are defined in Eq. (\ref{mij}).
 We note that to obtain
the unpolarized spectrum, we need to sum up the polarizations of
$\epsilon_{i}$ with $\sum_{\lambda} \epsilon^{*\mu}_{i}(\lambda)
\epsilon^{\nu}_{i}(\lambda)=-g^{\mu \nu}+p_{i}^{\mu} p_{i}^{\nu}/
m^{2}_{\phi}$. 

\begin{figure}[phbt]
\includegraphics*[width=2.3in]{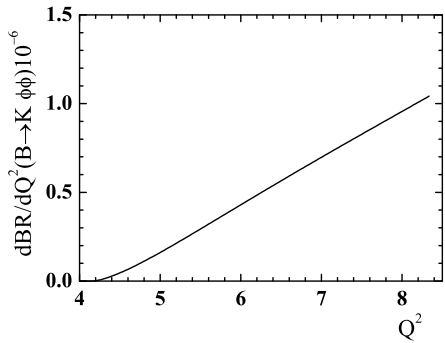} \caption{
The differential decay rates (in units of $10^{-6}$) for $B\to K
\phi \phi$ with the invariant mass of $\phi$ meson pairs below $2.85$
GeV.}
 \label{difrate}
\end{figure}

As known that the uncertain parts for the calculations of
exclusive decays are the hadronic matrix elements, such as the
functions of $Q^2$, $f_{\pm}$, $A$, $B_{1(2)}$, $C$, $D$ and $F$.
Since the form factors for $B\to K $ have been studied well in the
literature \cite{CG-NPB,Ali}, their $Q^2$-dependent functions
could be
controlled with definite errors. For convenience, according to the
results of Ref.~\cite{CG-NPB}, we parametrize the form factors
$f_{\pm}(Q^2)$ to be
\begin{eqnarray}
f_{+}(Q^2)&=& 0.35 \left(1-1.246\left(\frac{Q^2}{m^{2}_{B}}\right)+0.251\left(\frac{Q^2}{m^{2}_{B}}\right)^2\right)^{-1},\nonumber\\
f_{-}(Q^2)&=&0.35
\left(1-0.297\left(\frac{Q^2}{m^{2}_{B}}\right)-0.40\left(\frac{Q^2}{m^{2}_{B}}\right)^2\right)^{-1}.
\end{eqnarray}
 Moreover, since the remaining time-like form factors for $\langle \phi
\phi | V_{\mu} (A_{\mu})|0\rangle$ are not studied yet, to
get numerical estimations, we assume that they all the time-like form
factors have the same magnitude, {\it i.e.}, $B_1\sim B_{2}\sim
C\sim D\sim F$. In the following, we use ${\cal F}(Q^2)$ to denote
these form factors. In order to express the form
factor as a function of $Q^2$, we adopt the following form
\begin{eqnarray}
{\cal F}(Q^2)=e^{i\delta}\left(
\frac{a}{Q^2}-\frac{b}{Q^4}\right)\left[\ln \frac{Q^2}{d^2}
\right]^{-1},
\end{eqnarray}
where $\delta$ represents the strong phase. The expansion of
$(1/Q^2)^n$ is inspired from  Ref. \cite{Chua} for the $\langle K
K^* | V_{\mu} (A_{\mu})|0\rangle$ transition and the factor
$1/\ln(Q^2/d^2)$ is due to the clue of perturbative QCD
\cite{BF}. Since  the BR of $B\to K \phi \phi$ has been measured by Belle, we can use the experimental data to fit the
unknown parameters $a$, $b$ and
$d$. With the fitted parameters, we can
estimate the CP and T violating effects in $B\to K \phi \phi$
decays. Hence, in the SM with B$(B\to K \phi
\phi)_{Q<2.85 GeV}=2.0\times 10^{-6}$, we set $a=5.$, $b=4.$ and
$d=1.0$. The spectrum of the differential decay rate is shown in
Fig.~\ref{difrate}. Our figure is consistent with that of Ref.
\cite{FPP} in which the authors dressed the problem by considering
all possible intermediate states.

As emphasized early that to study T violating effects, we
have to investigate the polarizations of $\phi$ mesons. Since
$\phi$  decays to $KK$ dominantly, we expect that the T
violating terms could be related to the angular distribution of
$K_1$ and $K_2$, in which $K_{1}$ denotes the daughter of one of
two $\phi$ mesons while $K_{2}$ is that of the other $\phi$ meson.
The four-component momenta of $K_1$ and $K_2$ in their present
rest frame are chosen as follows:
$p_1=(E_1,E_1\sin\theta_1,0,E_1\cos\theta_1)$ and
$p_2=(E_2,E_2\sin\theta_2\sin\phi,E_2\sin\theta_2
\cos\phi,E_2\cos\theta_2)$ with $E_1=E_2=m_{\phi}/2$. We note that
$E_{1(2)}\approx |\vec{p}_{1(2)}|$
due to the smallness of the kaon mass.
Here, $\theta_{1(2)}$ are the
polar angles of K-mesons in each $\phi$ meson rest frame. The
angle $\phi$ represents the relative angle between two decaying
planes, produced by the two $\phi$-meson decays. Hence, the
angular distribution associated with T odd terms in
$B\to K\phi\phi$ is
obtained as
\begin{eqnarray}
{d\Gamma_{T-odd}(\theta_{1},\theta_2,\phi,Q^2) \over dQ^2
d\cos\theta_{1} d\cos\theta_{2}d\phi}&=&\frac{9}{4} {G^{2}_{F}
\over 2^{11} \pi^4  m_{B}}B^{2}(\phi\to KK)
\left(1-\frac{Q^2}{m^{2}_B}\right)\sqrt{1-\frac{(2m_{\phi})^2}{Q^2}}\nonumber \\
&& \times \left\{ - \left[\frac{1}{4} \int_{-1}^{1}
\rm{Im}(H_{0}(H^*_{-}-H^*_{+}))d\cos\theta \right] \sin2\theta_{1}
\sin2\theta_{2} \sin\phi \right.\nonumber \\
          &&\left.+\left[\frac{1}{2} \int_{-1}^{1}
\rm{Im}(H_{+}H^*_{-})d\cos\theta \right]\sin^2\theta_{1}
\sin^2\theta_{2} \sin2\phi  \right\},  \label{tv} \\
H_{0}&=&H(0,0)=p_{B}\cdot(p_1-p_2)\frac{Q^2}{(2m_{\phi})^2}
\nonumber \\ && \times\left[ 2m_{1}\left(1-\frac{2m^2_{\phi}}{Q^2}
\right)+m_{2}\left(1-\frac{(2m_{\phi})^2}{Q^2} \right)\right],
\nonumber \\
                  H_{\pm}&=&H(\pm,\pm)=-m_{1}p_{B}\cdot(p_1-p_2)\mp 2 m_{4}
|\vec{p}_{\phi}|E_{B} \mp m_{5} |\vec{p}_{\phi}|
\sqrt{Q^2},\nonumber
\end{eqnarray}
where $H_0$ and $H_{\pm}$ are the longitudinal and transverse
polarizations, respectively, and $B(\phi\to KK)$ is the decay branching
ratio of $\phi\to KK$. Clearly, the T odd terms
are related to not only angles $\theta_{1(2)}$ but also the
azimuthal angle $\phi$. We note that the results do not depend on
the angle $\theta$, which represents the polar angle of the $\phi$
meson in the $Q^2$ rest frame. To study these effects, we define
the statistical significances by \cite{CG-NPB}
\begin{eqnarray}
\bar{\varepsilon}_i={\int {\cal
O}_{i}\omega_{i}(u_{\theta_{K_1}},u_{\theta_{K_2}})d\Gamma \over
\sqrt{\int d\Gamma \cdot \int {\cal O}_i^2d\Gamma }} \label{op}
\end{eqnarray}
where $\omega_{i}(u_{\theta_{K_1}},u_{\theta_{K_2}})
=u_{\theta_{K_1}}u_{\theta_{K_2}}/
|u_{\theta_{K_1}}\,u_{\theta_{K_2}}| $ are sign functions with
$u_{\theta_i}$ being $\cos \theta_i$ or $\sin \theta_i$. In the
$Q^2$ rest frame, the T odd momentum correlations for operators in
Eq. (\ref{op}) are given by
\begin{eqnarray*}
{\cal O}_{T_{1}}&=& | \vec{p}_{B}| \frac{\vec{p}_{K_1}\cdot (
\vec{ p}_{B}\times \vec{p}_{K_2}) }{| \vec{p}_{B}\times \vec{p}
_{K_1}| | \vec{p}_{B}\times \vec{p}_{K_2}| } =\sin
\phi,\label{OT1}
\\
{\cal O}_{T_{2}}&=& | \vec{p}_{B}| \frac{( \vec{p}_{B}\cdot \vec{
p}_{K_2}\times \vec{p}_{K_1}) ( \vec{p}_{B}\times \vec{p}%
_{K_1}) \cdot ( \vec{p}_{K_2}\times \vec{p}_{B}) }{|
\vec{p}_{B}\times \vec{p}_{K_1}| ^{2}| \vec{p}_{K_2}\times \vec{p}%
_{B}| ^{2}} =\frac{1}{2}\sin 2\phi, \label{OT2}
\end{eqnarray*}
 accompanied with sign functions of $\omega_{T_{1}}(\cos
\theta_{K_1}, \cos \theta_{K_2})$ and $\omega_{T_{2}}(\sin
\theta_{K_1}, \sin \theta_{K_2})$, respectively.

Although Eq. (\ref{tv}) could indicate the T violating effects,
 since the definition in Eq. (\ref{op})
 does not represent the real time
reversal operator in which the initial state will be reversed to be
the final state, the appearance of strong phases also contributes
to Eq. (\ref{tv}). That is, $d\Gamma_{T-odd}\propto
\sin(\theta_W+\theta_s)$ where $\theta_W$ and $\theta_S$ are the
weak CP and strong phases, respectively. In order to avoid the
ambiguity for the nonvanished weak CP  and strong phases, we
propose to include the corresponding CP-conjugate mode and
define the new quantities as
\begin{eqnarray}
\bar{\varepsilon}_{i}(B)+\bar{\varepsilon}_{i}(\bar{B}) & \propto
& \sin(\theta_W+\theta_s)+\sin(-\theta_W+\theta_s)=2\cos\theta_W \sin\theta_s, \label{cp1} \\
          \bar{\varepsilon}_{i}(B)-\bar{\varepsilon}_{i}(\bar{B}) & \propto
& \sin(\theta_W+\theta_s)-\sin(-\theta_W+\theta_s)=2\sin\theta_W
\cos\theta_s. \label{cp2}
\end{eqnarray}
Evidently, if a nonvanished value of Eq. (\ref{cp1}) is observed ,
it will indicate the non-negligible relative strong phase between
time-like form factors. On the other hand, if  nonvanished value of Eq.
(\ref{cp2}) is measured, it will imply the existence of new
physical CP violating phase. Since our purpose is to probe the new
CP phases, we concentrate our discussions on the definition of
Eq.~(\ref{cp2}). The problem, whether the strong phases play
important contributions, is referred to the experiments.

\begin{figure}[phbt]
\includegraphics*[width=2.3in]{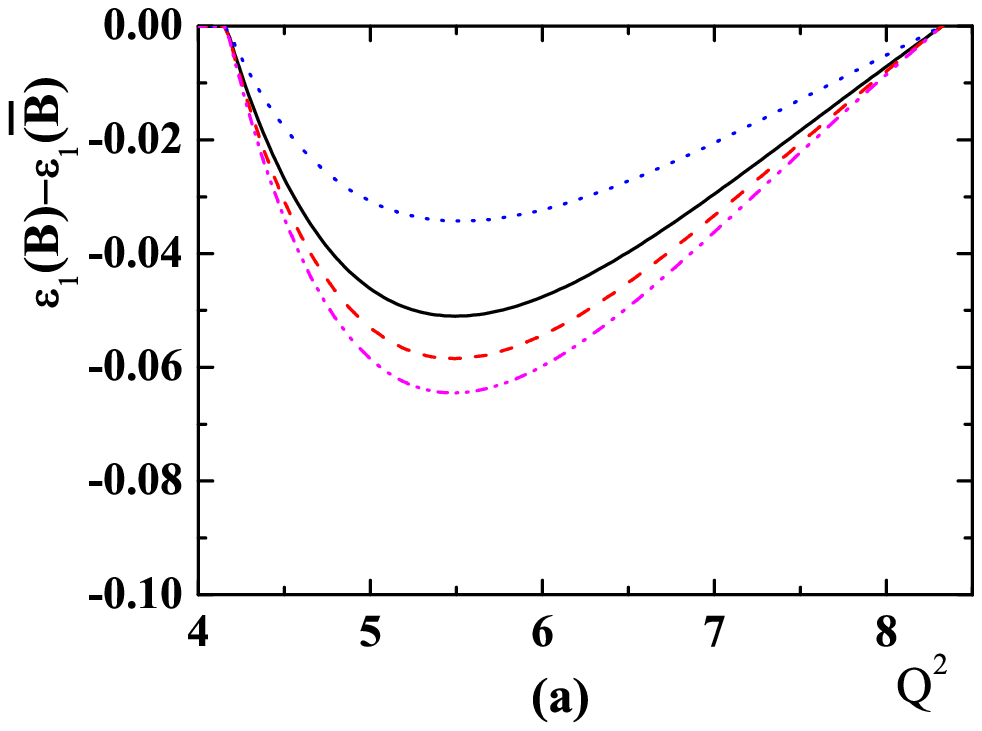} \hspace{0.5cm}\includegraphics*[width=2.3in]{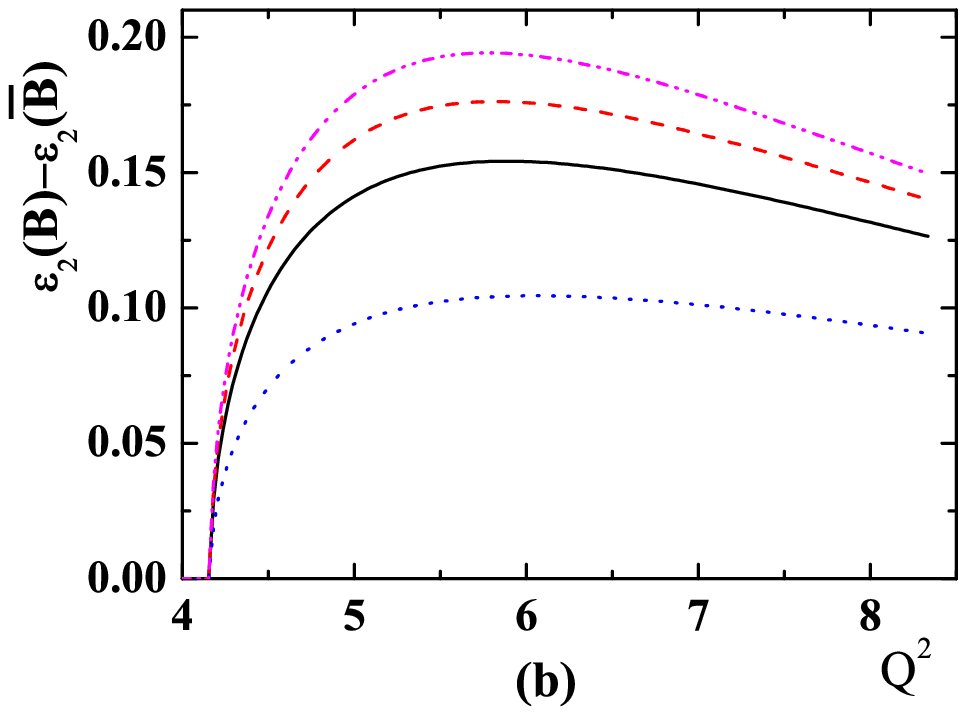}\caption{
The significances of T violation for (a) ${\cal O}_{T_1}$ and (b)
${\cal O}_{T_2}$ with respect to the invariant mass of $\phi$
meson pairs below $2.85$ GeV. The sold, dashed, dotted,
dash-dotted line stand for the different vales of $c^{NP}_{i}$. The
detailed description is in the text.}
 \label{figcp}
\end{figure}

To
illustrate the possibility of observing T violation at $B$ factories, 
instead of discussing a specific model,
we consider a generic class of CP violating
new physics interactions with
$c^{eff}_{k}=c^{SM}_{k}+e^{i\theta_{k}} |c^{NP}_{k}|$, where
$(c^{SM}_1,\,c^{SM}_2,\,c^{SM}_{3})=(-0.043,0.033,-0.053)$ are the values in the SM while
$\theta_k$ and $c^{NP}_k$ are related to new physics. For
simplicity, we take all $\theta_{k}=\pi/2$. In Fig. \ref{figcp},
we present the significances of T violation for some different
values of $|c^{NP}_i|$. The solid, dashed, dotted, dash-dotted
lines stand for
$(c^{NP}_1,\,c^{NP}_2,\,c^{NP}_3)=(-0.02,-0.04,-0.03)$,
$(-0.01,-0.05,-0.03)$, $(-0.06,-0.02,-0.04)$, and
$(0,-0.06,-0.03)$, with the
corresponding BRs
in turn being $2.53,\, 2.63,\, 2.78$, and $2.77\times 10^{-6}$,
respectively. According to the results of Fig. \ref{figcp}, we
clearly see that the contribution of ${\cal O}_{T_2}$ is much
larger than that of ${\cal O}_{T_1}$; and the effect could be more
than 10\%. We note that to measure this T violating effect at
$2\sigma$ level, at least $1.5\times 10^8$ B decays are required
if we use B$(B\to K\phi\phi)=2.6\times 10^{-6}$. Certainly, it
could be detectable at the $B$ factories.

Finally, we give some remarks on the resonant contributions
to the decays.
It was pointed out in
Ref. \cite{FPP} that the main resonant contributions to the decay
BR are
from $\eta_{c} (2980)$ and the changes are around
$\pm 10\%$, depending on the constructive or destructive
interference. Although the width of $\eta_{c}$ is as small as 
$17.3$ MeV, since the spectrum for the decaying rate is
increasing at $Q^{2}\sim 2.85$, as shown in Fig. \ref{difrate},
 the influence on the decay BR may not be neglected.
However, 
the
T-odd effects as shown in Fig. \ref{figcp} are decreasing when $Q^2$ is approaching to the
upper limits of data. In order to avoid the contributions from
resonant effects, we can search the T-odd effects in the region
which is far away from the resonant state $\eta_{c}$. In our
study, the best searching region of $Q^2$ is between $5$ and
$7$ GeV$^2$.

In summary, by the factorization assumption, we have studied the
T-odd observables in $B\to K \phi \phi$ decays. Despite the
hadronic uncertainties, we find that the T violating
effect  for ${\cal O}_{T_2}$ could reach $10 \%$. Although the
resultant depends on the strong phases, as shown in Eq.
(\ref{cp1}) and (\ref{cp2}), we can define the proper T-odd
observables associated with the CP conjugate modes so that the
experiments can tell us how much the effects are from the CP
conserved strong phases.\\

{\bf Acknowledgments}\\

This work is supported in part by the National Science Council of
R.O.C. under Grant \#s: NSC-91-2112-M-001-053,
NSC-93-2112-M-006-010 and NSC-93-2112-M-007-014.


\begin{thebibliography}{99}

\bibitem{CCFT}  J.H. Christenson, J.W. Cronin, V.L. Fitch and R. Turly,
Phys. Rev. Lett. {\bf 13}, 138 (1964).

\bibitem{BelleCP}Belle Collaboration, K. Abe {\it et al.},
 Phys. Rev. D{\bf 66}, 071102 (2002).

\bibitem{BabarCP}Babar Collaboration, B. Aubert {\it et al.},
Phys. Rev. Lett.{\bf 89}, 201802 (2002).

\bibitem{CKM}  N. Cabibbo, Phys. Rev. Lett. {\bf 10}, 531 (1963); M.
Kobayashi and T. Maskawa, Prog. Theor. Phys. {\bf 49}, 652 (1973).

\bibitem{BVV1} G. Valencia, Phys. Rev. D{\bf 39}, 3339 (1989).

\bibitem{BVV2}
A. Datta and D. London, Int. J. Mod. Phys. A{\bf 19}, 2505 (2004).

\bibitem{Todd} BABAR Collaboration, J.G. Smith, hep-ex/0406063, contribution to Moriond QCD proceedings;
 BELLE Collaboration, K. Abe {\it  et al.}, hep-ex/0408141.

\bibitem{CG-PRD66} C.H. Chen and C.Q. Geng, Phys. Rev. D{\bf 66}, 014007
(2002).

\bibitem{Belle-PRL} Belle Collaboration, H.C. Huang {\it et al.}, Phys. Rev. Lett. {\bf 91}, 241802
(2003).

\bibitem{Hazumi} M. Hazumi, Phys. Lett. B{\bf 583}, 285 (2004).

\bibitem{CG-NPB} C.H. Chen and C.Q. Geng,
Nucl. Phys. B{\bf 636}, 338 (2002); C.H. Chen and C.Q. Geng, Phys.
Rev. D{\bf 66}, 014007 (2002).

\bibitem{Ali} A. Ali {\it et al.}, Phys. Rev. D{\bf 61}, 074024 (2000)
; D. Melikhov and B. Stech, Phys. Rev. D{\bf 62}, 014006 (2000).

\bibitem{Chua}C.K. Chua {\it et al.}, Phys. Rev. D{\bf 70}, 034032 (2004).

\bibitem{BF}S.J. Brodsky and G.R. Farrar, Phys. Rev. D{\bf 11},
1309 (1975).

\bibitem{FPP} S. Fajfer, T.N. Pham, and A. Prapotnik, Phys. Rev.
D{\bf 69}, 114020 (2004).

\end{thebibliography}
\end{document}